\documentclass[conference]{IEEEtran}
\usepackage{cite}
\usepackage{amsmath,amssymb,amsfonts}
\usepackage{algorithmic}
\usepackage{cite}
\usepackage{amsmath,amssymb,amsfonts}
\usepackage{algorithmic}
\usepackage{graphicx}
\usepackage{textcomp}
\usepackage{xcolor}
\usepackage{float}
\usepackage{longtable}
\usepackage{tabulary}
\usepackage{tabularx}
\usepackage{dblfloatfix}
\usepackage{outlines}
\usepackage{placeins}
\usepackage{booktabs}
\usepackage{color, colortbl}

\usepackage[flushleft]{threeparttable}

\usepackage[english]{babel}
\usepackage{fancyhdr}
\usepackage{lastpage}

\usepackage{booktabs}

\def\BibTeX{{\rm B\kern-.05em{\sc i\kern-.025em b}\kern-.08em
    T\kern-.1667em\lower.7ex\hbox{E}\kern-.125emX}}

\usepackage{enumitem}

\begin{document}
\bstctlcite{IEEEexample:BSTcontrol}

\title{Network Intrusion Detection System in a Light Bulb}

\author{
\IEEEauthorblockN{
Liam Daly Manocchio\IEEEauthorrefmark{1},
Siamak Layeghy\IEEEauthorrefmark{2},
Marius Portmann\IEEEauthorrefmark{3}
}
\IEEEauthorblockA{
\textit{School of Information Technology and Electrical Engineering} \\
\textit{University of Queensland}\\
Brisbane, QLD 4072, Australia\\
\IEEEauthorrefmark{1}liam@riftcs.com, \IEEEauthorrefmark{2}siamak.layeghy@uq.net.au,
\IEEEauthorrefmark{3}marius@ieee.org
}
}

\maketitle

\begin{abstract}
Internet of Things (IoT) devices are progressively being utilised in a variety of edge applications to monitor and control home and industry infrastructure.
Due to the limited compute and energy resources, active security protections are usually minimal in many IoT devices.
This has created a critical security challenge that has attracted  researchers' attention in the field of network security.
Despite a large number of proposed Network Intrusion Detection Systems (NIDSs), there is limited research into practical IoT implementations, and to the best of our knowledge, no edge-based NIDS has been demonstrated to operate on common low-power chipsets found in the majority of IoT devices, such as the ESP8266.
This research aims to address this gap by pushing the boundaries on low-power Machine Learning (ML) based NIDSs.
We propose and develop an efficient and low-power ML-based NIDS, and demonstrate its applicability for IoT edge applications by running it on a typical smart light bulb.
We also evaluate our system against other proposed edge-based NIDSs and show that our model has a higher detection performance, and is significantly faster and smaller, and therefore more applicable to a wider range of IoT edge devices.

\end{abstract}


%%%%%%%%%%%%%%%%%%%%%%%%%%%%%%%%%%%%%%%%%%%%%%%%%%%%%

\section{Introduction}

Internet of Things (IoT) edge devices are finding increasing use and prevalence in powerful device ecosystems, ranging from smart homes to remote sensor networks. There are also industrial scale IoT systems (IIoT) which have significantly higher levels of complexity than ordinary IoT Networks. It is estimated that there are over 14 billion IoT endpoints in 2022 \cite{Hasan2022NumberGlobally}.
Because of their widespread usage, and their applications in commercial industrial infrastructure, they have become the target of various cyberattacks.

Despite the fact that IoT devices are used to monitor and control `things' from home security systems, through to medical monitoring devices and industrial infrastructure, they often do not have the same level of protection that can be achieved on servers and workstations.
This is to a large extent due to their limited compute and energy resources, and their application in a diverse range of networks that make it more difficult to implement cybersecurity controls.

While there are many documented cases of compromise of IoT edge devices, including the incredibly damaging Mirai botnet in 2016 that compromised over 600K edge and embedded devices \cite{Antonakakis2017UnderstandingBotnet}, the average consumer does not have access to high grade network intrusion detection systems (NIDSs) that could be used to detect and protect against these types of attacks.
An edge-based NIDS can enable a fast reaction to attacks against IoT devices and networks, without needing to centrally process data. This decentralised processing also improves privacy, by allowing data to be kept local, at the edge of the network.

This security risk has not gone unnoticed in the research community, and there are works proposing several IoT compatible NIDSs, which we discuss in this paper.
However, many of these proposals are theoretical and do not evaluate their models on actual edge hardware. Some works were also tested on relatively high-power edge devices, such as Google's Edge TPU or Raspberry Pi \cite{Hosseininoorbin2021ExploringIoT}, which are unlikely to be widely deployed in typical  IoT devices and smart home networks.
A number of frameworks exist with the aim of bringing machine learning to the edge, such as TensorFlow Lite~\cite{Abadi2016TensorFlow:Systems}. However, NIDS models built using these frameworks are often still inaccessible to low-power microcontrollers \cite{Idrissi2022AIoT}.
The models in the literature we surveyed, based on TensorFlow Lite, have a large memory footprint, leaving little room for other functionality on typical low-end IoT devices.

\begin{figure}
    \centering
    \includegraphics[width=0.7\linewidth]{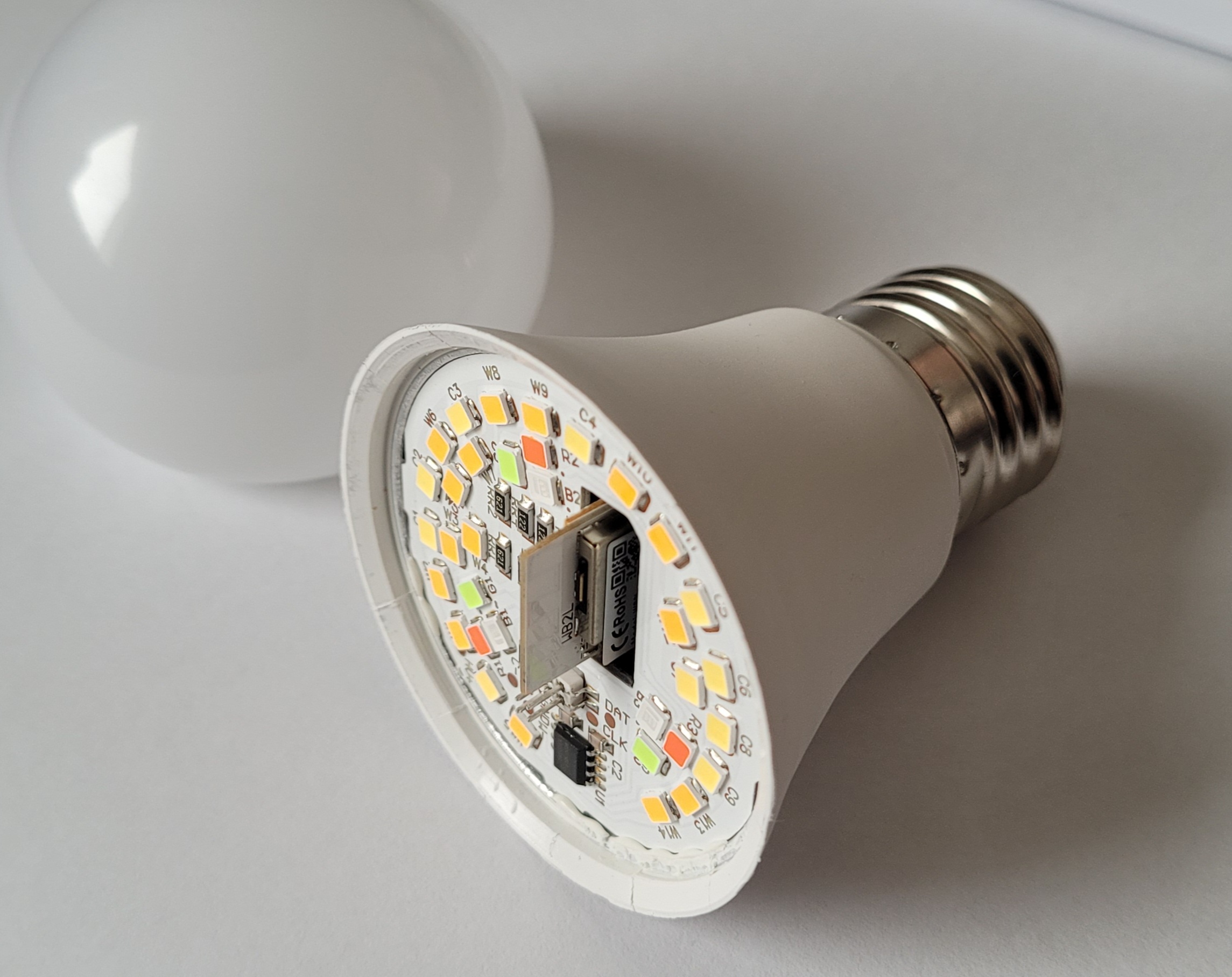}
    \caption{A photo of a typical consumer grade `Tuya' compatible smart light bulb that we use for our NIDS light bulb demonstration\protect\footnotemark}
    \label{fig:avglightbulb}
\end{figure}

\footnotetext{This is a picture of a newer model that uses the Tuya WB2L SoC, we use an earlier version with the ESP8266 microcontroller}

\begin{figure}[t]
    \centering
    \includegraphics[width=0.9\linewidth]{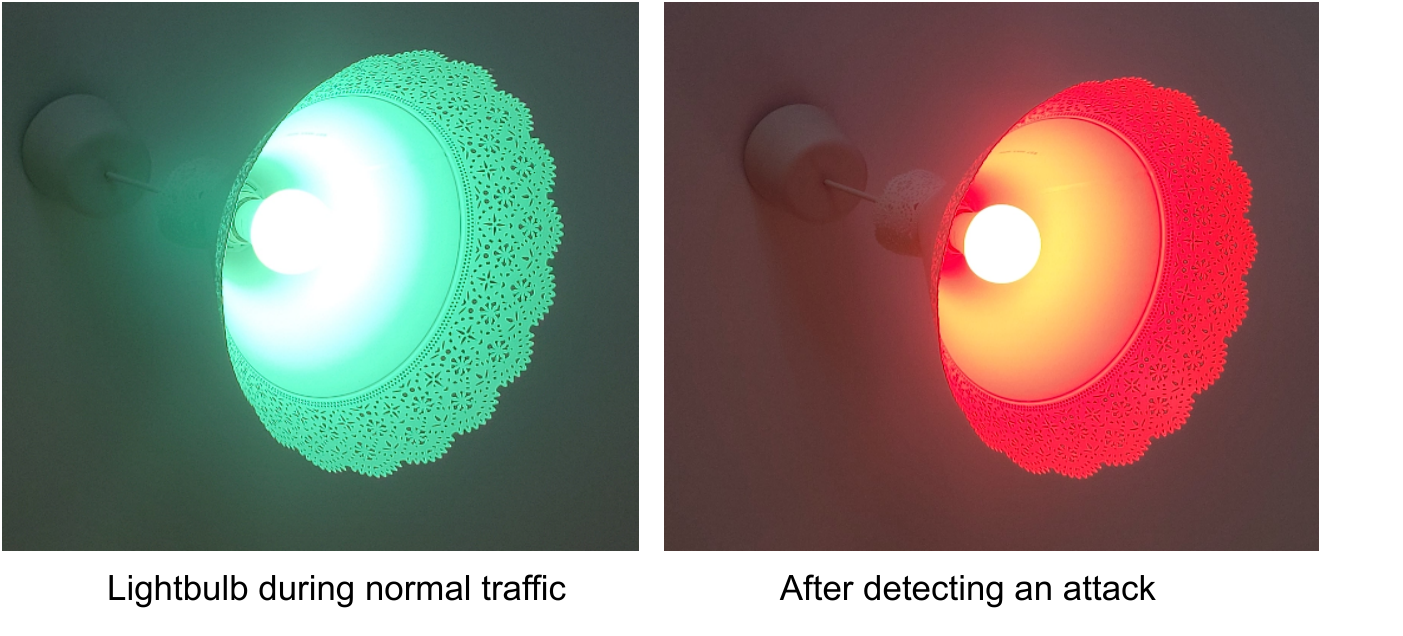}
    \caption{An example of our light bulb NIDS running on an ESP8266 microcontroller on a modified consumer smart light bulb. Green indicates normal traffic, whereas red indicates an attack has been detected.}
    \label{fig:demonstration}
\end{figure}

To address this gap, this research aims to push the bounds on what is possible in a low compute power environment, and to demonstrate that high accuracy network intrusion detection can be brought to the wider IoT domain. We propose, develop and evaluate a high performance NIDS capable of running on the lowest power conventional edge microprocessors.
We show that the performance of this NIDS is comparable to the existing approaches proposed in the literature, while being significantly faster and more lightweight.
To further demonstrate the applicability of the proposed NIDS at the IoT edge, we deploy it on a typical smart light bulb, and demonstrate the world's first NIDS in a light bulb.
A similar smart light bulb is shown in Figure~\ref{fig:avglightbulb}. To do this, we replace the ESP8266 microcontroller on a consumer smart light bulb, with a transplant ESP8266 microcontroller featuring our modified NIDS firmware. 
Then as a fun way to demonstrate our NIDS running on the smart bulb, we control the colour of the light emitted by the bulb, i.e. green during normal operation, and red when an attack is detected, shown in Figure~\ref{fig:demonstration}. 

The key contribution of this paper is the proposal, implementation, and evaluation of an extremely lightweight NIDS, capable of functioning on the lowest-power IoT devices. We have made the code publicly available here \footnote{Code available at https://rft.io/lightbulb}.
Based on our experimental evaluation, our system outperforms the state-of-the-art IoT NIDS proposals on IoT hardware both in terms of detection accuracy, detection speed and resource requirements. 

\section{Related Works}

There have been several efforts to develop lightweight and scalable machine learning systems for use in IoT devices and networks. 
There are many previous works that adopt a signature-based approach for intrusion detection. 
However, signature based detection suffers from the limitation that signatures must be manually updated. Since this paper focuses on ML-based NIDSs, these works have not been included in our discussion.

In terms of machine learning based NIDS, these can be grouped into shallow learning and deep learning. The authors in \cite{Chaabouni2019NetworkTechniques} and \cite{Tabassum2019AIoTs} surveyed several approaches to NIDSs in IoT devices, and found a variety of works that used shallow learning to great success.
For example, \cite{Alhowaide2021TowardsNIDS} evaluated five classifiers following feature selection, PCA based anomaly detection, a local deep SVM, a logistic regression and a boosted decision tree. Across three benchmark datasets, the authors showed 100\% accuracy for all approaches other than PCA.
There are also several proposed approaches in the literature that utilise deep learning models.
For instance, \cite{Hodo2017ThreatSystem} uses a multi-layer perceptron (MLP) model, which is a fully connected dense artificial neural network, to achieve 99.4\% accuracy. There are also several deep unsupervised approaches, such as \cite{Lopez-Martin2017ConditionalIoT}, which showed that using an autoencoder, a sufficiently low reconstruction loss could be achieved for networking data, to facilitate an IoT compatible anomaly detection system. More advanced forms of neural networks, such as graph neural networks, have also been proposed for IoT devices \cite{Lo2022E-GraphSAGE:IoT}, and these have achieved F1 scores of 0.81 on two IoT benchmark datasets.
However, these systems discussed here that have achieved 99\%+ accuracy were not tested on real IoT hardware.

There has been a limited number of works that have tested proposed NIDSs on real IoT hardware.
The use of Google's Edge TPU platform has been explored for use with NIDS models \cite{Hosseininoorbin2021ExploringIoT}. Here, the authors compared the performance of a convolutional neural network (CNN) running on a Google Edge TPU with that of a Raspberry Pi (Cortex-A53), and demonstrated fast performance as well as 0.98+ F1 scores. However, both Edge TPU and Raspberry Pi have significantly more processing power than the average IoT smart device.
\cite{Idrissi2022AIoT} is the most relevant to our work, since it implements a deep learning based NIDS on several lower power hardware platforms \cite{Idrissi2022AIoT}, including ESP32-WROOM-32, ESP8266 and ATmega328p.
The authors used TensorFlow Lite for their approach, which allowed them to bring a pre-trained neural network model to various microcontrollers \cite{Abadi2016TensorFlow:Systems}. They were able to achieve 96.7\% detection accuracy on the ESP32-WROOM-32. However, the proposed model was too large to be deployed on low-end devices such as ATMega328p.
Furthermore, the authors' ESP8266 implementation used nearly 100\% of the device's memory, making it impractical for parallel deployment to an existing low-end IoT device, where a significant amount of memory is likely required for the code and data of the devices' core functionality. 

There exist several solutions, outside of TensorFlow Lite, that allow machine learning models to be brought to microcontrollers. 
Of particular interest here are solutions that can convert models developed in scikit-learn \cite{Pedregosa2011Scikit-learn:Python}, another widely used machine learning framework, to microcontroller compatible code. These tools, which include sklearn-porter, and EmbML \cite{TsutsuiDaSilva2019EmbMLSystems}, are capable of porting pre-trained scikit-learn models to microcontrollers. However, unlike Tensorflow Lite which is primarily focused on deep learning models, scikit-learn features many shallow learning approaches.
To the best of our knowledge, no previous work has used these techniques to take pre-trained NIDS models and run them on IoT hardware. 

In summary, although there are many works that demonstrate high accuracy for IoT NIDSs when tested on benchmark data, there has been relatively limited experimental research into the practical implementation and deployment of NIDSs on IoT edge hardware.
The research that has been conducted to date delivers NIDSs that either require too much processing power to operate, or would utilise too much of device resources to be applicable as part of a smart device with inbuilt intrusion detection functionality.
Our work presented in this paper aims to address this gap by proposing, implementing and evaluating a high-performance ML-based NIDS, capable of running on resource-constrained edge IoT devices.

\section{Lightbulb NIDS}
\subsection{Datasets}

For training and evaluating the performance of ML-based NIDSs, data is required. For this, there are publicly available and highly cited benchmark datasets. These datasets are usually captures of network data generated synthetically or on test beds, designed for NIDS research. 
Since the main approaches to obtain network traffic include packet-capture (pcap) and flow-based traffic collection, publicly available NIDS datasets are often represented in one or both of these formats.
In the packet-based approach, the full packet headers and payloads are captured as they are sent across the network. 
In the flow-based approach, only aggregate information about the network traffic is collected, based on the sequence of packets between two endpoints. There are many formats of flow based data.

As a first step in the design of our proposed NIDS for edge IoT, it is important to decide on the format of the input data.
Continuous packet-based network monitoring is very resource intensive, and is typically not feasible for large scale networks, particularly large-scale IoT networks, where thousands of devices may communicate their status in a short time period.
Flow-based network traffic monitoring, on the other hand, is more scalable and includes less information and thus has fewer security and privacy issues. 
Accordingly, we chose the NetFlow as the data format in this study, which is also consistent with prior works in this space \cite{Idrissi2022AIoT}\cite{Hosseininoorbin2021ExploringIoT}.

This paper considers five different widely used and highly cited NIDS datasets. 
\begin{enumerate}[align=right,itemindent=2em,labelsep=2pt,labelwidth=1em,leftmargin=0pt,nosep]
    \item \textit{Ton-IoT}, an IoT and industrial IoT dataset featuring `various attacking techniques, such as DoS, DDoS and ransomware, against web applications, IoT gateways and computer systems across the IoT/IIoT network' \cite{Moustafa2021ADatasets}
    \item \textit{BoT-IoT}, an IoT dataset featuring various botnet traffic, including `DDoS, DoS, OS and Service Scan, Keylogging and Data exfiltration attacks' \cite{Koroniotis2018TowardsDataset}
    \item \textit{MQTT-IoT-IDS2020} (MQTT), an IoT dataset where `five scenarios are recorded: normal operation, aggressive scan, UDP scan, Sparta, SSH brute-force, and MQTT brute-force attack.' \cite{Hindy2021MachineDataset}.
    \item \textit{UNSW-NB15}, an NIDS dataset for traditional networks featuring `a hybrid of real modern normal activities and synthetic contemporary attack behaviours' with `nine types of attacks, namely, Fuzzers, Analysis, Backdoors, DoS, Exploits, Generic, Reconnaissance, Shellcode and Worms' \cite{Moustafa2015UNSW-NB15:Set}
    \item \textit{CSE-CIC-IDS2018}, an NIDS dataset for traditional networks including `seven different attack scenarios: Brute-force, Heartbleed, Botnet, DoS, DDoS, Web attacks, and infiltration of the network from inside' \cite{Sharafaldin2018TowardCharacterization}
\end{enumerate}
These datasets which form the basis of our evaluation, represent a range of several classes of network traffic and attack types. These datasets are available in different feature sets.
For the BoT-IoT, ToN-IoT, UNSW-NB15 and CSE-CIC-IDS2018 datasets, we used the version of these converted to a standardised flow format which was proposed by \cite{Sarhan2021TowardsDatasets}. Works using these derivative datasets have shown comparable performance to works using the datasets in their original packet capture format.
For the MQTT dataset, we use the published bidirectional flow format data released with the original capture.

\subsection{Model Choice}
Finding a suitable machine learning model, which can satisfy the requirements of an efficient and lightweight NIDS at the IoT edge, is critical.
While \textit{Deep Learning (DL)} models have been shown to be very successful in the implementation of ML-based NIDSs in general networks~\cite{Idrissi2022AIoT}, they usually need a large amount of memory and compute resources.
As such, we considered several \textit{Shallow Learning} approaches that are known to be less resource intensive, and we decided to use a \textit{decision tree} model based on its generally low complexity and high classification performance. 

This model choice is supported by the literature, several related works demonstrate that tree-based models can be effectively used for NIDSs, including \cite{Uhm2021Service-AwareScalability}, \cite{Alhowaide2021TowardsNIDS} and \cite{Chew2020DecisionSystem}.
In~\cite{Uhm2021Service-AwareScalability} a 95.25\% detection accuracy with a random forest on the Kyoto 2016 dataset is achieved. In \cite{Alhowaide2021TowardsNIDS} an accuracy of 100\% with three benchmark datasets is achieved using a boosted tree. 
Finally, in \cite{Chew2020DecisionSystem} a decision tree with sensitive pruning is applied to the GureKDDCup dataset to achieve 99\%+ accuracy. 
In addition to these previous works, three other factors motivated our use of decision trees:

\begin{enumerate}[align=right,itemindent=2em,labelsep=2pt,labelwidth=1em,leftmargin=0pt,nosep]
    \item The results of our early experimentation showed comparable accuracy of tree based models to that of larger machine learning models, such as random forests~\cite{Uhm2021Service-AwareScalability} and even neural networks~\cite{Idrissi2022AIoT}.
    \item It can be expected that the number of instructions required to implement a decision tree would be far fewer than other types of shallow machine learning models such as random forests, while also balancing predictive power versus `too simple' models such as linear support vector classifiers.
    \item Decision trees have decision paths that can be easily analysed for model interpretability. They also translate logically to code that can be easily read and interpreted by programmers. This property was noted in \cite{Chew2020DecisionSystem} as an advantage.
\end{enumerate}

Despite the use of tree based models in previous works, we were unable to find an implementation of these models on actual IoT hardware.
We therefore aimed to investigate the implementation of decision trees, with the goal of balancing low complexity and resource requirements with high detection performance, both in terms of accuracy and speed.

\subsection{Pre-processing}

\begin{figure}[t!]
    \centering
    \includegraphics[width=0.6\linewidth]{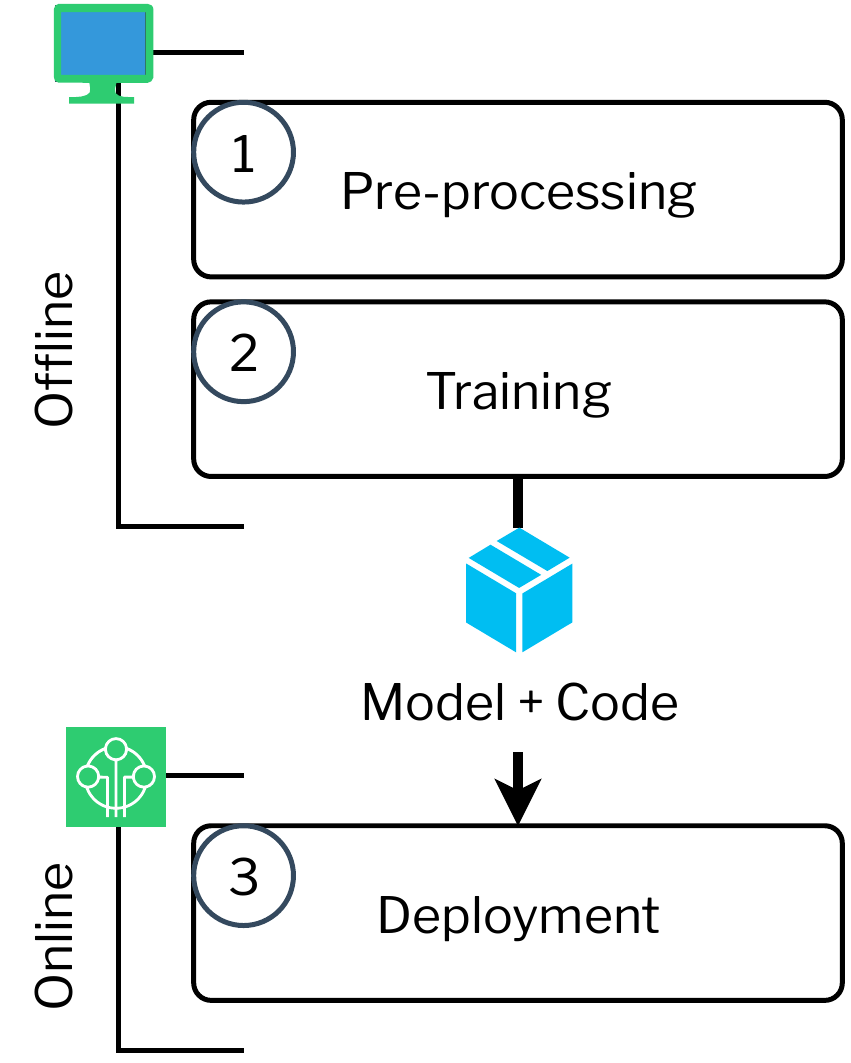}
    \caption{Our approach begins with the offline  pre-processing of data and training the machine learning model, followed by the deployment of this pre-trained model to an IoT device which can ingest flow data in real-time.}
    \label{fig:our_approach_summarised}
\end{figure}

Figure~\ref{fig:our_approach_summarised} shows the basic stages of our approach. This involves training a machine learning model using benchmark datasets offline, and then testing this pre-trained model on real IoT edge hardware.
As can be seen, the first stage in this pipeline is pre-processing, which involves transforming data into a format that is suitable for machine learning models. Decisions regarding pre-processing must take into account the data format, as well as the type of model being used.

As previously discussed, for the purposes of this work, we ingest data in flow-based format. 
Despite this format being a very common and useful representation for network administrators, it typically requires some processing to make it suitable for use in a machine learning model.
Network data in flow format has several fields that are categorical. For example, TCP and UDP port numbers. Although ports 22 (SSH) and 25 (SMTP) are close numerically, they represent vastly different protocols. Whereas port 8080 and port 443 are both commonly used for HTTPS despite the larger numerical difference. Certain machine learning models, such as neural networks, typically function best when the distances between values represent some contextual distance. There are several standard approaches to pre-processing that can be applied to categorical data to solve these issues, such as ordinal, frequency, or target encoding.
In addition to the categorical fields, there are several numerical fields in flow based data, such as the number of bytes or packets. Numerical data is often standardised prior to use in machine learning, to ensures that the inter-feature variance is equal. This can help in the training of certain classes of machine learning models.

However, as discussed, pre-processing must also be done with consideration of the model being trained. Because tree based models work by splitting data and since this can be done at an arbitrary point, regardless of scale, they are somewhat robust to data that has not been pre-processed both in terms of categorical data or unscaled numerical data. This is in contrast to neural networks that typically require significant pre-processing.
We tested tree based models with and without pre-processing and found that the performance was comparable, these results are shown in Table~\ref{tab:preprocessing-performance}. In later investigations, we found that pre-processing also increased the inference time.
Because of this, we opted to not perform pre-processing. As the flow format data we are using expresses all features as numeric data types, these can be directly handled by decision tree models.

In order to ensure stable model training offline, we did ensure that training data had a balanced representation of the benign and attack class. Training on imbalanced data is a significant challenge for many models, and we solve this by using random under sampling to collect a balanced training dataset that is smaller than the overall dataset. However, we perform a separate cross validated evaluation step to ensure that all samples are considered when performing evaluation, using metrics that are resistant to imbalance such as balanced accuracy or F1 score.

\begin{table}
\caption{Performance of a low max-depth decision tree model, with and without pre-processing}
\centering
\begin{tabular}{@{}lcc@{}}
\toprule
\textbf{\begin{tabular}[c]{@{}l@{}}Dataset \\ (Balanced Accuracy)\end{tabular}} & \textbf{\begin{tabular}[c]{@{}c@{}}With\\ Pre-processing\end{tabular}} & \textbf{\begin{tabular}[c]{@{}c@{}}No\\ Pre-processing\end{tabular}} \\ \midrule
\multicolumn{1}{c}{\textbf{ToN-IoT}}                                 & 98.75\%                                                                & 98.70\%                                                              \\
\multicolumn{1}{c}{\textbf{BoT-IoT}}                                                       & 98.70\%                                                                & 98.70\%                                                              \\ \bottomrule
\end{tabular}
\label{tab:preprocessing-performance}
\end{table}

\begin{figure}[!b]
    \centering
    \includegraphics[width=1\linewidth]{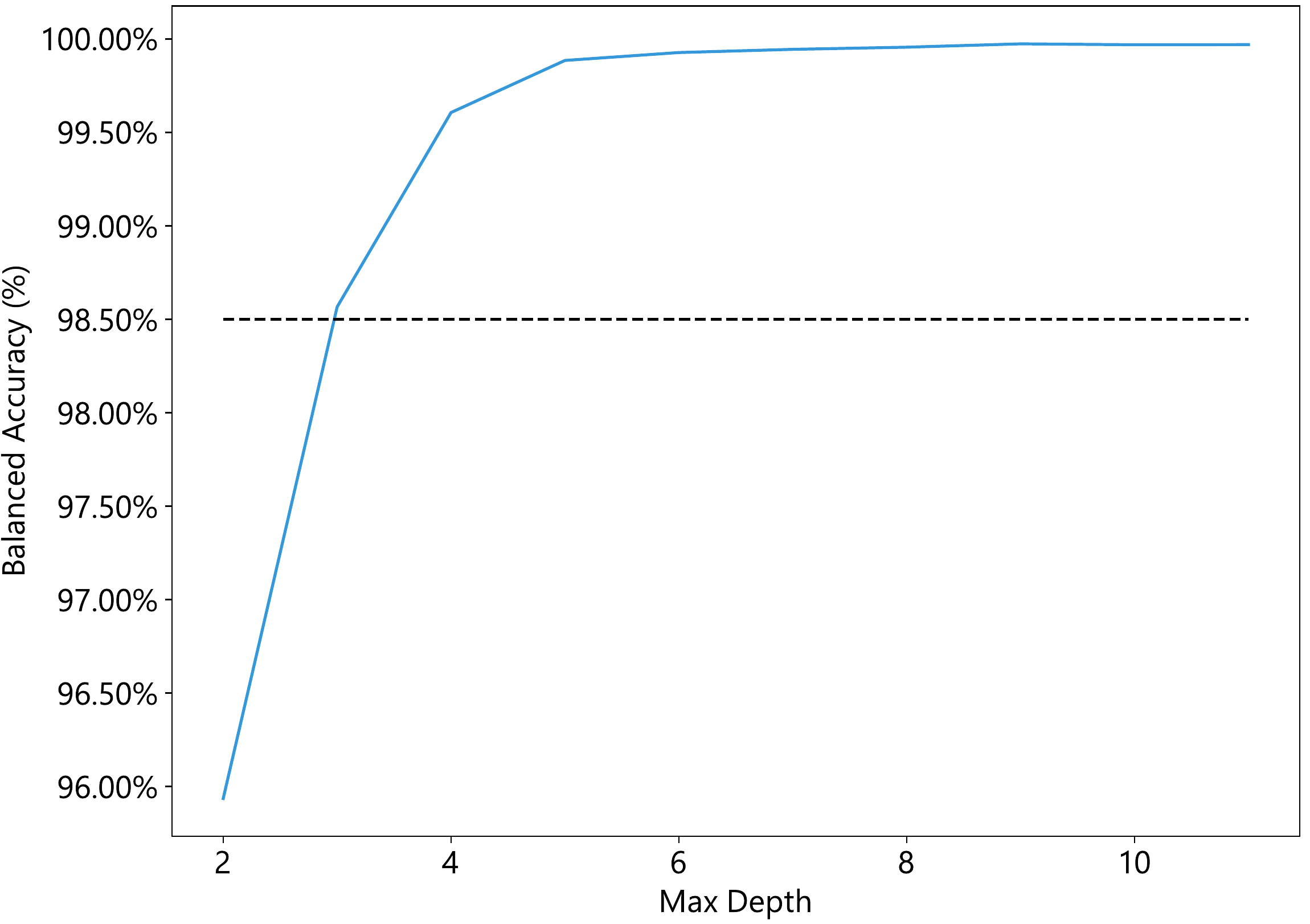}
    \caption{Averaged performance of various decision tree depths when trained on the BoT-IoT training split, versus their balanced accuracy on the holdout data split. We can see that models with a depth greater than 6 all converge to an accuracy of near 100\% performance. The line at 98.5\% shows our initial acceptance criteria, so we can rule out depths below 4.}
    \label{fig:depth_versus_f1}
\end{figure}

\subsection{Training and Model Hyperparameters}

\begin{table}[t!]
    \caption{Hyperparameters chosen for our decision tree model}
    \centering
    \includegraphics[width=0.55\linewidth]{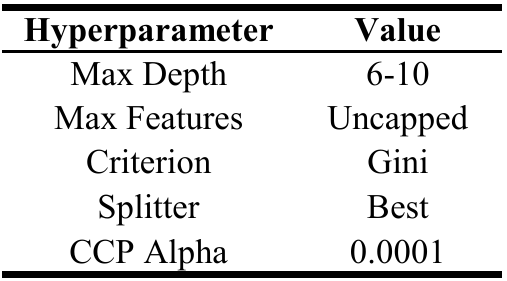}
    \label{tab:hyperparameter_choices}
\end{table}

To train the models, we used the Python language and the scikit-learn library for machine learning models. Scikit-learn is a commonly used machine learning library in scientific research, and is an open source tool accessible to machine learning researchers \cite{Pedregosa2011Scikit-learn:Python}.

Within scikit-learn, we used the Decision Tree Classifier. The hyperparameters of our decision tree are shown in Table~\ref{tab:hyperparameter_choices}; they represent the default values used in the used library, except for the ones specifically discussed below.
There are several parameters that can be varied when using decision trees, but the most important one is the max depth of the tree. 
We performed an initial systematic search of the max-depth parameter space between a max-depth of 2 and 12, on the BoT IoT dataset, to determine which candidate depths had an acceptable balanced accuracy. Here we used traditional holdout evaluation rather than cross validation. The results of this are shown in Figure~\ref{fig:depth_versus_f1}.
This allowed us to pick 6 and 10 as max-depth targets for investigation, 6 being the lowest depth with a reasonable accuracy of $>$98.5\% and 10 at the point of diminishing return where all later depths were near 100\% accuracy. We also tested depth 5 and depth 12 which lie slightly outside this range.

Our second non-standard parameter choice is the \textit{Cost Complexity Pruning (CCP) Alpha}, which is only relevant if cost complexity pruning is used. 
Cost Complexity Pruning is a common method used to address overfitting, and has also been shown experimentally to produce significantly smaller decision trees with similar detection accuracy \cite{Bradford1998PruningCosts}. 
This was confirmed during our initial investigation, in which we found that using a CCP Alpha value of 0.0001 reduced the number of features and model size, without diminishing the accuracy significantly.

One of the other common parameter choices is the \textit{Max Features} parameter, which is used during the initial tree fitting to limit the number of features to split on. However, our tests indicated that leaving this uncapped yielded the best results. 
We also used the \textit{Gini index}, with a splitting strategy of best, which is the default for the scikit-learn library. 
The Gini index can be thought of as the purity of a given set of observations, with 0 indicating all observations belong to a single class, and 1 indicating a random distribution between all classes.
This is used to determine how to split a tree as part of the classification and regression trees algorithm (CART), which chooses the feature to split on at each node to achieve the best or highest purity. 
This was achieved in earlier decision tree algorithms such as ID3 or C4.5 by maximising the information gain at each node. Scikit-learn \cite{Pedregosa2011Scikit-learn:Python} utilises an optimised version of the CART algorithm.

\subsection{Transferring a Trained Model to a Microcontroller}
To convert our models to C-code for use by a microcontroller, we developed a source-to-source converter in Python that accepts a scikit-learn tree model, and converts the tree into C code.
\begin{figure}[t!]
    \centering
    \includegraphics[width=0.9\linewidth]{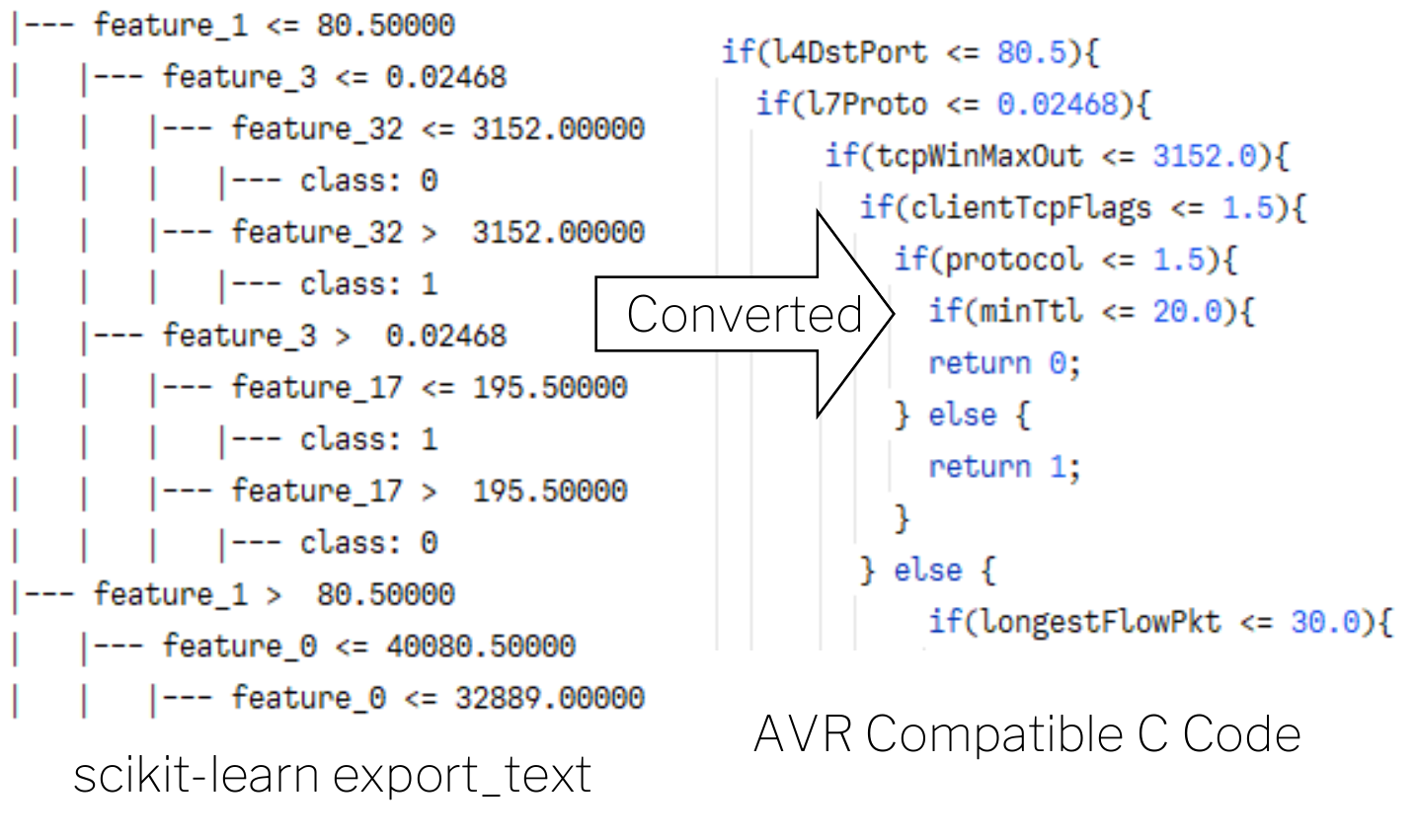}
    \caption{Our source-to-source translator accepts a scikit-learn decision tree, and uses the output from the scikit-learn export\_text function to convert this into a C decision tree.}
    \label{fig:transpilation}
\end{figure}
This C code is then injected into a template that can be complied for a target microcontroller. This template includes the functionality to read the flow data from the serial port, as well as the ability to time the model. An example of this conversion is shown in Figure~\ref{fig:transpilation}.
Our approach is different to those used in other scikit-learn frameworks, as we utilise nested if statements to express decision trees.
Other approaches utilise recursion for each depth as well as a global array of thresholds, but we expect that our approach utilising nesting will compile to a smaller number of instructions, when used for smaller decision trees such as those in our proposed NIDS. 

\section{Experimental Methodology}

In this section, we discuss how we evaluate our proposed approach. 

\subsection{Microcontroller}

\begin{table}[b!]
    \caption{Characteristics of the three devices used in this work.}
    \centering
    \includegraphics[width=1\linewidth]{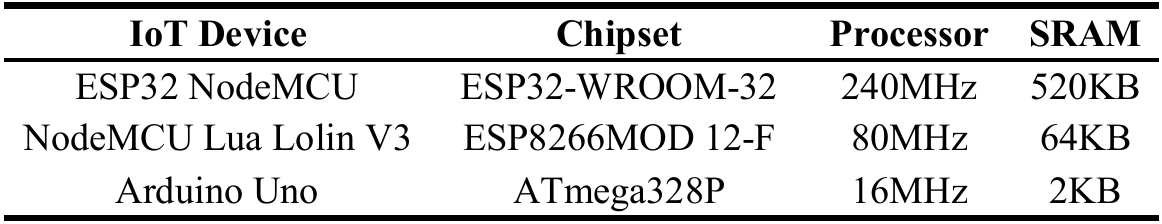}
    \label{tab:microcontroller_comparison}
\end{table}

We evaluate our work against three low-power microcontrollers, shown in Table~\ref{tab:microcontroller_comparison}. We refer to the ESP32 Node MCU as ESP32, the NodeMCU Lua Lolin V3 as ESP8266, and the Arduino Uno as ATMega328p, based on their chipset.
These represent three of the lowest-power IoT devices in widespread use, with the ESP8266 having found widespread applications in all manners of smart devices, from smart light bulbs to entry control systems. The ESP32 is a slightly more powerful device, but still finds use in higher-end IoT devices.
The ATMega328p on the other hand is the chip powering several Arduino products. Arduino has found widespread use in the hobbyist space. Despite its low power, there exists several shields that can enable the Arduino Uno to connect to the internet via Ethernet or Wi-Fi.
We chose these devices because they allow us to compare our system's performance in the case of typical IoT devices. 

\subsection{Measuring inference time}

To measure inference time was relatively challenging, given that the onboard timer cannot necessarily be expected to have a reliable microsecond resolution. 
\begin{figure}
    \centering
    \includegraphics[width=0.8\linewidth]{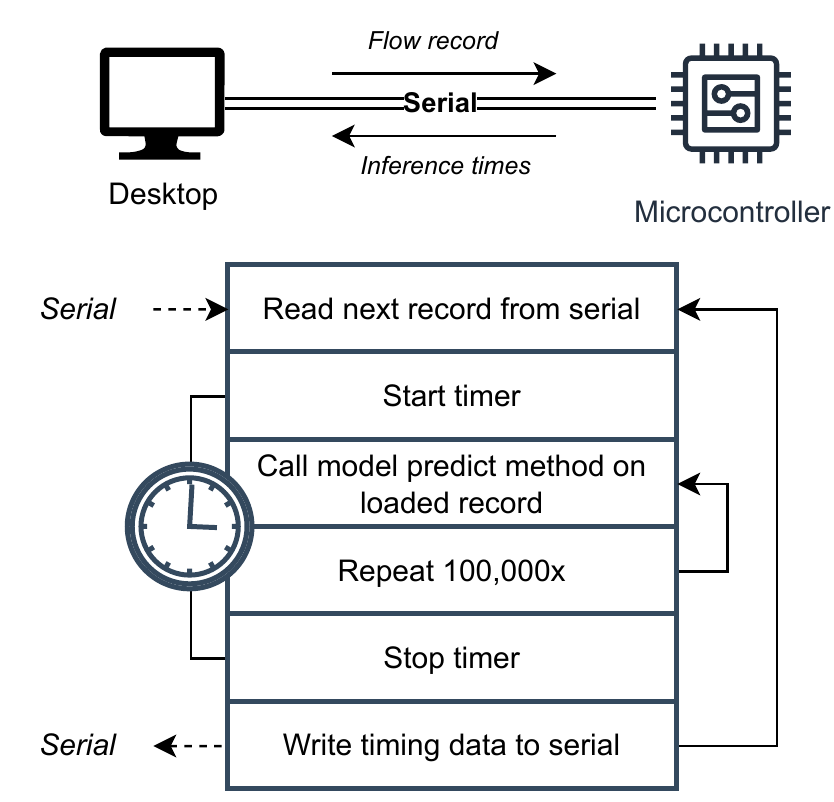}
    \caption{Our setup uses the onboard device clock to measure the inference time. To do this, we send flow records via serial and convert this into numerical features on the device. We then call the predict method on these features 100,000 times, timing how long this takes, before sending this timing data to the computer and proceeding to the next record. }
    \label{fig:inference_time}
\end{figure}
In addition, the `micros()' clock returns an integer, which does not allow us to measure fractions of microseconds.
To adjust for this inaccuracy, as can be seen in Figure~\ref{fig:inference_time}, we repeated calls to the predict function 100,000 times for each flow record, and then divided the total time taken for this to get the average inference time per record. 
Here, caution needs to be taken to ensure that the inference method is not optimised out during compilation, and that the results from repeated function calls are not cached at the CPU level during execution. 

In our experiment, we send flow records to the device via the serial port for convenience. In order to not distort the inference time measurement, we do not consider the transfer time. 

In a practical deployment of our NIDS in an IoT device, flow records would be generated in real-time from the network traffic observed on the Wi-Fi interface, and exported locally on the IoT device before being considered by the ML-based traffic classifier. 
However, because we are using benchmark datasets for testing, we instead send these records directly to the device.

\subsection{Measuring Accuracy}

For accuracy measurement, we used the same model parameters as when testing inference time, however, we apply repeated stratified cross validation with 5 splits and 5 repeats, measuring the balanced accuracy and F1 score of each of these splits.
This splits the dataset into 5 different balanced groups, and computes the validation accuracy holding out each of these groups in turn, averaging this, and then averaging this across 5 repeats of other random groupings.
This provides the most robust and reliable accuracy data, which is more likely to detect model overfitting. 
Because cross validation does not by default yield a single trained model that can be deployed, for timing results, we instead split data into a training and testing dataset, using standard holdout evaluation for accuracy. However, these scores are all in the 99.5\%+ range even for model depths of 5, as can be seen with the performances found in our initial investigation in Figure~\ref{fig:depth_versus_f1}.

\section{Results}

\begin{table}
    \caption{Our solution when compared to sklearn-porter for translating the same depth 12 decision tree to C code}
  \begin{threeparttable}
    \centering
    \includegraphics[width=1\linewidth]{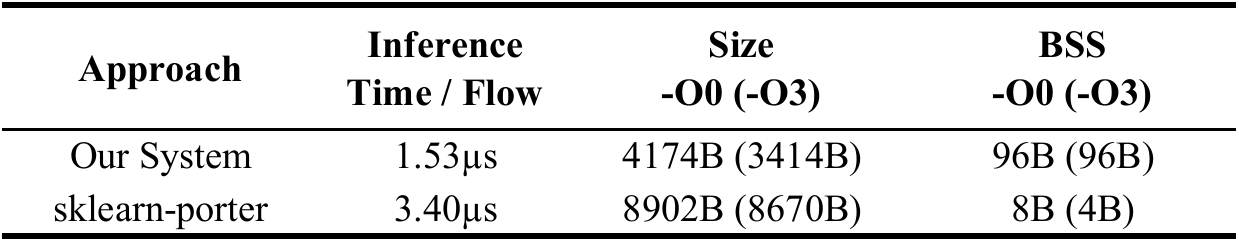}
     \begin{tablenotes}
      \item \begin{footnotesize}
          Size is measured with the Linux tool `size' given after compilation with GCC at optimisation level 0 (-O0) and 3 (-O3) respectively. Here, BSS (block starting symbol) size can be considered the size of global variables.
      \end{footnotesize}
    \end{tablenotes}
    \end{threeparttable}
     \label{tab:porter_comparison}
\end{table}

\begin{table}[t!]
    \caption{The average inference time per flow for varying model depths}
    \centering
    \includegraphics[width=1\linewidth]{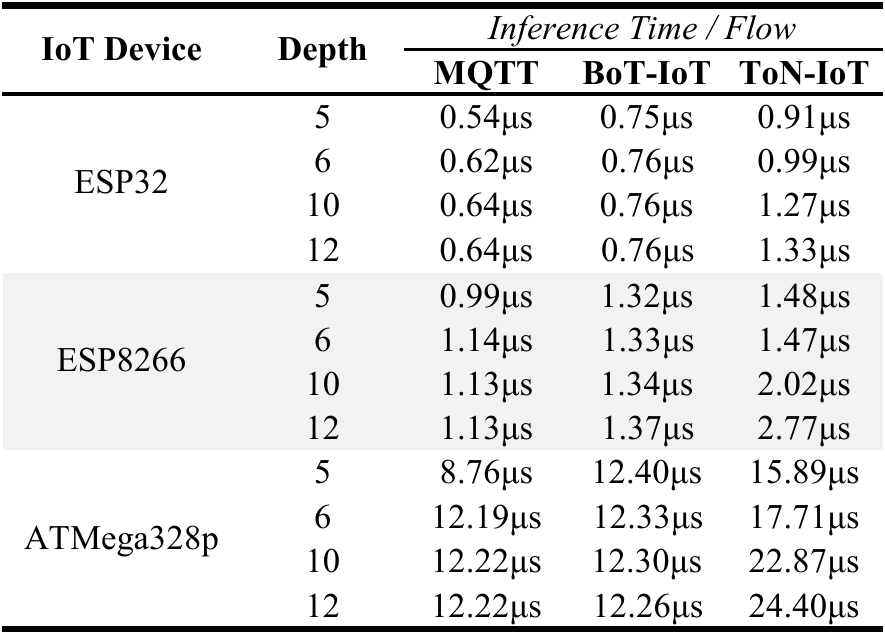}
    \label{tab:OurResultsTable_DeviceSpeed_Transposed}
\end{table}

\begin{table*}[t!]
    \caption{The balanced accuracy (BAcc.) and F1 scores (F1 is expressed as a percent for convenience) of the cross validated results on both traditional and IoT datasets for various model max depths. Best result is highlighted in bold.}
    \centering
    \includegraphics[width=1\linewidth]{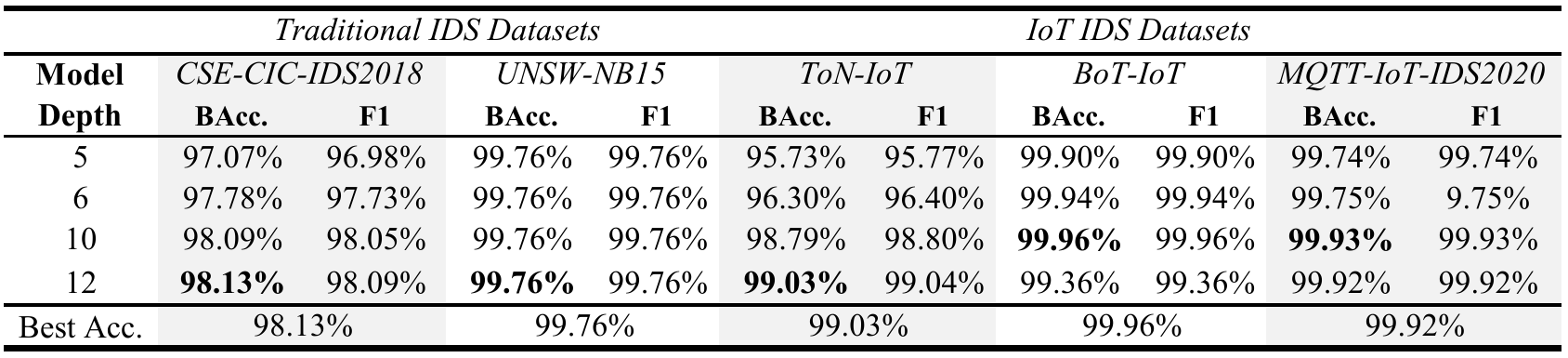}
    \label{tab:OurResultsTable_AccuraciesVsDataset}
\end{table*}

\begin{table*}[!b]
\caption{The inference time and performance of our depth 12 model versus other models in the literature}
    \centering
    \includegraphics[width=1\linewidth]{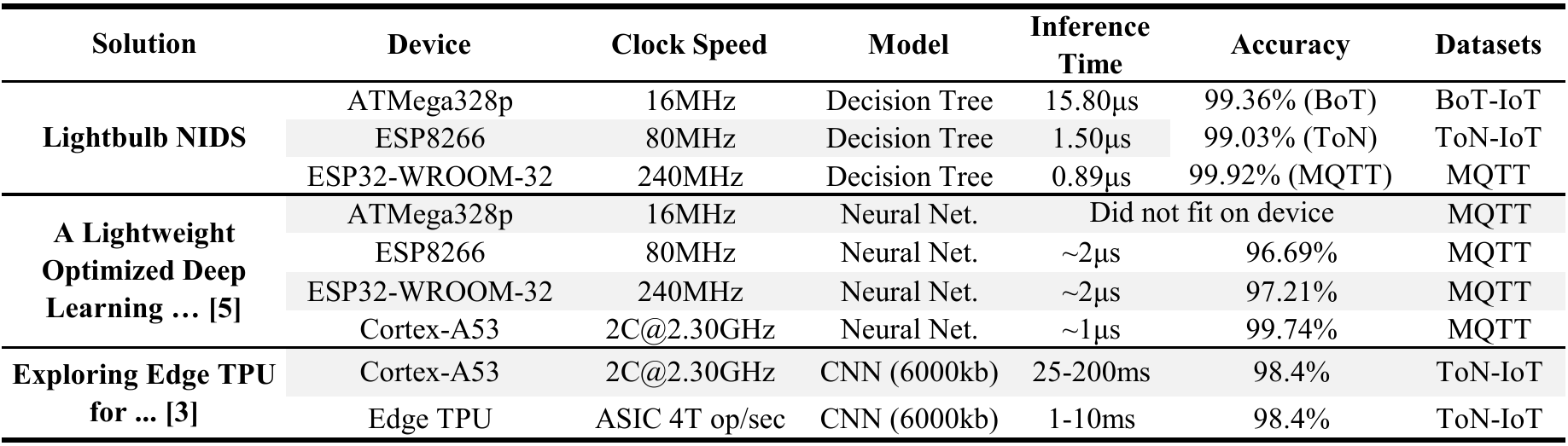}
    \label{tab:OurResultsTable_VersusLiterature}
\end{table*}

\subsection{Comparison to other Source-to-source Converters}

We begin by evaluating our choice to write a custom source-to-source converter for transforming the scikit-learn decision tree to a format that can be used on a microcontroller.
As discussed in the literature review, other tools do exist for the conversion of scikit-learn machine learning models to C-code. However, we hypothesised that a lean small decision tree focused approach could yield more optimised code compared to a general library. 
To test this, we compared our code transformer to that of sklearn-porter in terms of program size as well as inference time. The results are shown in Table~\ref{tab:porter_comparison}.

Here we compared the same depth 12 decision tree model, with the same bare-bones boilerplate code, substituting only the predict method with either the predict method generated by sklearn-porter or by our converter.
We used the GCC compiler with the `-O0` option to compare un-optimised program sizes, as well as `-O3` to compare sizes after optimisation. 
Timing data is from the deployment of this version of the model to the ESP32 with the ToN-IoT dataset at depth 12, although Table~\ref{tab:porter_comparison} shows a different trained model than in Table~\ref{tab:OurResultsTable_DeviceSpeed_Transposed}.
We can see from Table~\ref{tab:porter_comparison} that our program has a significantly smaller memory footprint than sklearn-porter, as well as a lower inference time.
We believe this is due to the logic used for the source-to-source conversion. For our solution with a relatively low decision tree depth, our nested `if-then' logic produces relatively simple and inexpensive chains of comparison instructions.
In contrast, sklearn-porter uses four integer arrays to store child nodes, as well as a large double array for thresholds. This contributes to a larger memory footprint. In addition, it uses recursive iteration for each layer of the tree, and at each layer of the tree iterates over the entire class array. This produces more condensed code in terms of number of lines, but may require additional operations for smaller trees compared to our approach.

However, our approach did have a higher Block Starting Symbol (BSS) which indicates we have more global variables.
This is because we initialise a set of global variables for each feature in a flow record, and process the incoming flows directly into these variables to remove the need for parameter passing.
This is still however well under 10\% of the dynamic memory of the ATMega328p, which is one of the lowest-power practical microcontrollers, so this is not a significant factor.

\subsection{Detection Accuracy and Performance}
Table~\ref{tab:OurResultsTable_DeviceSpeed_Transposed} shows the inference time per flow for our model, for each of the three considered benchmark datasets. We can see that CPU speed is reflected in inference time, as is the depth, although the impact of depth is minor. Finally, the results are roughly consistent between datasets, with ToN-IoT having the longest inference times.
Table~\ref{tab:OurResultsTable_AccuraciesVsDataset} shows 
the cross validated detection accuracy in terms of balanced accuracy and model scores for our model, for each of the five datasets. We can see that for most datasets the accuracy improves with model depth, but this is only a small improvement, and that for a depth 12 model, 99\%+ accuracy is achieved for all three IoT datasets, and 98\%+ for the traditional network NIDS datasets.
Based on the accuracy and inference time results, we decided to compare our model at depth twelve, to models in the related works.

Table~\ref{tab:OurResultsTable_VersusLiterature} shows these key metrics compared with two related works. 
We give the average inference time of our depth 12 model across all three IoT datasets as per Table~\ref{tab:OurResultsTable_DeviceSpeed_Transposed}. We also use accuracy results from our depth 12 model as per Table~\ref{tab:OurResultsTable_AccuraciesVsDataset}. We only consider the IoT specific datasets for comparison.
Our model achieved a higher accuracy than other approaches, even than those on more powerful hardware. Additionally, the results of our experiments show that our decision tree was able to function at a significantly faster speed than the related works. 
For example, our model on the ESP32 had an inference time of 0.89 microseconds, whereas \cite{Idrissi2022AIoT} took almost twice as long. Also, the accuracy of our model was 99.92\% on the MQTT dataset, compared to 97.21\% of \cite{Idrissi2022AIoT} on the same dataset. Our solution on the ESP32 is even faster than models from \cite{Idrissi2022AIoT} and \cite{Hosseininoorbin2021ExploringIoT} which ran on the significantly more powerful Raspberry Pi. 
In addition, \cite{Idrissi2022AIoT} was unable to fit a machine learning model on the Atmega328p, and the ESP8266 model had a lower accuracy due to the significant model compression used. Our uncompressed model was able to fit on both these devices with 99\%+ accuracy, and a sub-millisecond inference time on the ATMega328p.

\section{Conclusion}

In this paper, we present a highly efficient machine learning based NIDS that is deployable on extremely resource constrained IoT devices, such as a typical smart light bulb. Our experimental evaluation shows that our system outperforms the relevant state-of-the-art IoT NIDS solutions both in terms of detection accuracy and speed. 
Another key benefit of our solution is its extremely low memory footprint. 
For example, the size of our model at export only occupies around 10\% of the program space of the highly popular ESP8266 microcontroller, compared to the near 90\% utilisation of a comparable TensorFlow Lite model \cite{Idrissi2022AIoT}. This means that our model can be deployed on low-end IoT devices as an add-on service via a software upgrade. This opens up new possibilities to provide enhanced low-cost security services for IoT networks, which are under increasing threats of cyberattacks.

%% === END ===

\bibliographystyle{IEEEtran}
\bibliography{\jobname, references}

\end{document}